\documentclass[aps,superscriptaddress,11pt]{revtex4}
\usepackage{amsmath,amsfonts,amssymb,amsbsy,graphicx,bbm,hyperref,subfigure}

\begin{document}
\title{One-way Quantum Deficit and Decoherence for Two-qubit $X$ States}

\author{Biao-Liang Ye}
\affiliation{School of Mathematical Sciences, Capital Normal University,
Beijing 100048, China}
\author{Yao-Kun Wang}
\affiliation{Institute of Physics, Chinese Academy of Sciences,
Beijing 100190, China}
\affiliation{College of Mathematics,  Tonghua Normal University,
Tonghua, Jilin 134001, China}
\author{Shao-Ming Fei}
\affiliation{School of Mathematical Sciences, Capital Normal University,
Beijing 100048, China}
\affiliation{Max-Planck-Institute for Mathematics in the Sciences,
Leipzig 04103, Germany}

\begin{abstract}
We study one-way quantum deficit of two-qubit $X$ states systematically from analytical derivations.
An effective approach to compute one-way quantum deficit of two-qubit $X$ states has been provided.
Analytical results are presented as for detailed examples.
Moreover, we demonstrate the decoherence of one-way quantum deficit under phase damping channel.

\end{abstract}


\maketitle

\section{Introduction}
Quantum entanglement is one of the most distinguishing properties in quantum mechanics,
which gives quantum information processing novel advantages over classical information processing \cite{hhhh,horodecki09}.
Recently, quantum correlations \cite{modi12} receive much attention because they may play
vital roles in quantum information processing and quantum
simulation even without quantum entanglement \cite{datta08,lanyon08}. The characterization and quantification
of quantum correlations have become one of the significant topics in the past decade \cite{modi12}.
One of the quantum correlations is characterized by
quantum discord \cite{zurek01,henderson01} which is shown to play important roles in
quantum information tasks such as quantum state
discrimination \cite{roa11,libo12}, remote state preparation \cite{da12} and
quantum state merging \cite{maca,cavalcanti11}.
There have been many kinds of quantum correlations like measurement-induced disturbance \cite{luo2},
geometric quantum discord \cite{daluo,luo10}, relative entropy of
discord \cite{modi}, continuous-variable discord \cite{adesso, giorda} etc..
However, analytical computation of these quantum correlations seem extremely difficult
as optimization involved. Few analytical results have been
obtained even for general two-qubit states. An analytical formula of
quantum discord for Bell-diagonal states is provided in \cite{luo08}.
For general two-qubit $X$ states, the analytical formula is still
missed \cite{ali10,lu11,chen11,davi,huang13}. Recently,
the authors in \cite{loa,yur,yur15} presented a better classification in deriving analytical
quantum discord for five parameters $X$ states.

Among other quantum correlations, the quantum deficit is related to extract work from a correlated system coupled to a heat bath
under nonlocal operations \cite{opp02}, with the work deficit defined by $W_{t}-W_{l}$,
where $W_{t}$ is the information of the whole system and $W_{l}$ is the localizable information \cite{horo05}.
It is also equal to the difference of the mutual information and the classical deficit \cite{oppenheim2}.
The analytical formula of one-way quantum deficit \cite{opp02,horo05}, like quantum discord, remains unknown.
In Ref.\cite{wyk} the quantum deficit of four-parameter two-qubit $X$ states has been calculated.
Numerical results for five-parameter $X$ states are presented in Ref.\cite{wyk14}. In this paper,
by using different approaches, we systematically compute the one-way quantum deficit for general
two-qubit $X$ states in terms of analytical derivations. Analytical results are presented for classes of detailed quantum states.
Decoherence of one-way quantum deficit under phase damping channels is calculated too.

\section{One-way quantum deficit of two-qubit $X$ states}

We consider two-qubit states $\rho_{AB}$ in Hilbert space $\mathcal{H}_A^2\otimes \mathcal{H}_B^2$.
The one-way quantum deficit \cite{str11} is defined according to the minimal increase of
entropy after measurement on $B$,
\begin{eqnarray}\label{de}
  \tilde{\Delta}(\rho_{AB})=\min_{M_k} S(\sum_k M_k\rho_{AB} M_k)
  -S(\rho_{AB}),
\end{eqnarray}
where the minimum is taken over all measurement operators $\{M_k\}$ satisfying $\sum_k M_k=\mathbbm{1}$,
$S(\rho)=-{\rm Tr}\rho \log_2 \rho$ is von Neumann entropy.
It is equal to the thermal discord \cite{zurek03}. It is also denoted by the relative entropy to the set
of classical-quantum states \cite{horo05}.

Since the quantum correlations between $A$ and $B$ do not change under the local unitary operations,
we consider $\rho_{AB}$ in the Bloch representation as
\begin{eqnarray}\label{bloch}
  \rho_{AB}=\frac14[I\otimes I +x \sigma_3\otimes I
  + y I\otimes  \sigma_3+\sum_{i=1}^3t_i\sigma_i\otimes\sigma_i],
\end{eqnarray}
where $\sigma_i$ $(i=1,2,3)$ are the Pauli matrices,
$x,y,t_1,t_2$ and $t_3$ are real number.
Equivalently under the computational bases $\{|00\rangle, |01\rangle, |10\rangle, |11\rangle\}$,
\begin{eqnarray}\label{x}
 \rho_{AB}= \left(\begin{array}{cccc}
    a & 0 & 0 & f\cr
    0 & b & e & 0\cr
    0 & e & c & 0\cr
    f & 0 & 0 & d
  \end{array}\right),
\end{eqnarray}
where
\begin{eqnarray}
a&=&\frac{1}{4} \left(1+t_3+x+y\right);~~~
b=\frac{1}{4} \left(1-t_3+x-y\right);\\
c&=&\frac{1}{4} \left(1-t_3-x+y\right);~~~
d=\frac{1}{4} \left(1+t_3-x-y\right);\\
e&=&\frac{1}{4} \left(t_1+t_2\right);~~~~~~~~~~~~~
f=\frac{1}{4} \left(t_1-t_2\right).
\end{eqnarray}
$\rho_{AB}$ is called  two-qubit $X$ state, in which
the parameters satisfy the relations $a,b,c,d,e,f\ge0$, $a+b+c+d=1$, $|e|^2\le bc$ and $|f|^2\le ad$.

It is easily verified that $S(\rho_{AB})$ in (\ref{de})
is given by $S(\rho_{AB})=-\sum_\pm (u_\pm\log_2 u_\pm+ v_\pm\log_2 v_\pm)$, with
\begin{eqnarray}
  u_\pm&&=\frac14(1+t_3\pm\sqrt{(x+y)^2+(t_1-t_2)^2}),\nonumber\\[2mm]
  v_\pm&&=\frac14(1-t_3\pm\sqrt{(x-y)^2+(t_1+t_2)^2}).
\end{eqnarray}

For arbitrary rank-two two-qubit states, the projective
measurements are optimal to minimizes the von Neumann entropy \cite{shi}.
They are also almost sufficient rank-three and four two-qubit states \cite{galve11}.
In the following, we focus on projective measurements.
Let $M_k$ be the local measurement operators on subsystem $B$, $M_k=V\Pi_k V^{\dagger}$,
with $\Pi_k=|k\rangle\langle k|$, $k=0,1$, and $V\in U(2)$ unitary matrices.
Generally $V$ can be expressed as $V=t I+ i \vec{y}\cdot\vec{\sigma}$, where
$t\in R$ and $\vec{y}=(y_1, y_2, y_3)\in R^3$ satisfy $y_1^2+y_2^2+y_3^2+t^2=1$.
After measurement the state $\rho_{AB}$ is transformed to the ensemble $\{\rho_k, p_k\}$ with
\begin{equation}
\rho_k=\frac{1}{p_k}(I\otimes M_k)\,\rho_{AB}\,(I\otimes M_k),
\end{equation}
and $p_k={\rm tr}(I\otimes M_k)\rho_{AB}(I\otimes M_k)$.
By tedious calculation \cite{wyk14}, we have $\sum_k (I\otimes M_k)\,\rho_{AB}\,(I\otimes M_k)=p_0\rho_0+p_1\rho_1$, where
\begin{eqnarray}
  p_0\rho_0&=&\frac14[I+y z_3 I+ t_1 z_1 \sigma_1+ t_2 z_2\sigma_2
  + (x+ t_3 z_3)\sigma_3]\otimes V\Pi_0 V^\dagger,\\
  p_1\rho_1&=&\frac14[I-y z_3 I- t_1 z_1 \sigma_1- t_2 z_2\sigma_2
  - (x- t_3 z_3)\sigma_3]\otimes V\Pi_1 V^\dagger,
\end{eqnarray}
$z_1=2(-t y_2+ y_1 y_3)$, $z_2=2(ty_1+y_2 y_3)$, $z_3=t^2+y_3^2-y_1^2-y_2^2$.

From the matrix diagonalization techniques in Ref. \cite{wyk},
we have the eigenvalues of $\sum_k (I\otimes M_k)\,\rho_{AB}\,(I\otimes M_k)$,
\begin{eqnarray}
  \lambda_{1,2}=\frac14\left(1+y z_3\pm\sqrt{(x+ t_3z_3)^2+ t_1^2 z_1^2+ t_2^2 z_2^2}\right),\nonumber\\[2mm]
  \lambda_{3,4}=\frac14\left(1-y z_3\pm\sqrt{(x- t_3z_3)^2+ t_1^2 z_1^2+ t_2^2 z_2^2}\right),
\end{eqnarray}
with $z_1^2+z_2^2+z_3^2=1$. Therefore the one-way quantum deficit (\ref{de}) is given by
\begin{equation}\label{13}
\tilde{\Delta}=\sum_{\pm} (u_\pm\log_2 u_\pm+v_\pm \log_2 v_\pm)+\min [-\sum_{i=1}^4 \lambda_i\log_2\lambda_i].
\end{equation}

To find the analytical solutions of (\ref{13}), let us set $z_1=\sin\theta\cos\varphi$,
$z_2=\sin\theta\sin\varphi$ and  $z_3=\cos\theta$. Then
\begin{eqnarray}
\lambda_{1,2}=\frac{1}{4} \left(p_{+}\pm\sqrt{R+S_+}\right),~~~~
\lambda_{3,4}=\frac{1}{4} \left(p_{-}\pm\sqrt{R+S_-}\right),
\end{eqnarray}
where
$$
p_{\pm}=1\pm y \cos \theta,~~~\nonumber\\
R=[t_1^2 \cos ^2\varphi +t_2^2 \sin ^2\varphi ]\sin ^2\theta,~~~\nonumber\\
S_{\pm}=\left(x\pm t_3 \cos \theta \right)^2.
$$
Because $\lambda_i$ is the eigenvalue of the density matrix,
$\lambda_i\geqslant0$. Hence $p_{\pm}\geqslant\sqrt{R+S_{\pm}}\geqslant0$.

Denote $G(\theta, \varphi)=-\sum_{i=1}^4 \lambda_i\log_2\lambda_i$.
The one-way quantum deficit is given by the minimal value of $G(\theta, \varphi)$.
We observe that $G(\theta, \varphi)=G(\pi-\theta, \varphi)$ and
$G(\theta, \varphi)=G(\theta, 2\pi-\varphi)$. Moreover, $G(\theta, \varphi)$ is symmetric
with respect to $\theta=\pi/2$ and $\varphi=\pi$. Therefore,
we only need to consider the case of $\theta\in[0, \pi/2]$ and $\varphi\in[0,\pi)$.

The extreme points of $G(\theta, \varphi)$ are determined by
the first partial derivatives of $G$ with respect to $\theta$ and $\varphi$,
\begin{eqnarray}\label{cth}
\frac{\partial G}{\partial \theta}=-\frac{\sin\theta}{4}H_{\theta},
\end{eqnarray}
with
\begin{eqnarray}
H_{\theta}=&&\frac{R\csc\theta\cot\theta-t_3
\sqrt{S_+}}{\sqrt{R+S_+}}\log_2\frac{p_+ +\sqrt{R+S_+}}{p_+ -\sqrt{R+S_+}}+y\log_2\frac{p_-^2-(R+S_-)}{p_+^2-(R+S_+)}\nonumber\\[2mm]
&&+\frac{R\csc\theta\cot\theta+t_3
\sqrt{S_-}}{\sqrt{R+S_-}}\log_2\frac{p_- +\sqrt{R+S_-}}{p_- -\sqrt{R+S_-}},
\end{eqnarray}
and
\begin{eqnarray}\label{cph}
\frac{\partial G}{\partial \varphi}=2\,ef\sin^2\theta\,\sin2\varphi\, H_{\varphi},
\end{eqnarray}
with
\begin{eqnarray}
H_{\varphi}=\frac1{\sqrt{R+S_+}}\log_2\frac{p_+ +\sqrt{R+S_+}}{p_+ -\sqrt{R+S_+}}
+\frac1{\sqrt{R+S_-}}\log_2\frac{p_- +\sqrt{R+S_-}}{p_- -\sqrt{R+S_-}}.
\end{eqnarray}

Since $H_{\varphi}$ is always positive, $\frac{\partial G}{\partial \varphi}=0$ implies that
either $\sin2\varphi=0$, namely, $\varphi=0, \pi/2,$ for any $\theta$,
or $\theta=0$ for any $\varphi$ which implies that (\ref{cth}) is zero at the same time and
the minimization is independent on $\varphi$.
If $\theta\ne 0$, one gets the second derivative $\partial^2G/\partial\varphi^2$
\begin{eqnarray}
\frac{\partial^2G}{\partial\varphi^2}|_{(\theta,0)}
&&=4ef\sin ^2(\theta )H_{\varphi=0}>0,
\end{eqnarray}
and
\begin{eqnarray}
\frac{\partial^2G}{\partial\varphi^2}|_{(\theta,\pi/2)}
&&=-4ef\sin ^2(\theta )H_{\varphi=\pi/2}<0.
\end{eqnarray}
Since for any $\theta$ the second derivative $\partial^2 G/\partial\varphi^2$
is always negative for $\varphi=\pi/2$, we only need to deal with the minimization problem
for the case of $\varphi=0$. To minimize $G(\theta, \varphi)$ becomes to minimize $G(\theta, 0)$ which can be written as
$G(\theta, 0)=-\sum_{j,k=1}^2 w_{j,k}\log_2 w_{j,k}$ with
\begin{eqnarray}
  w_{j,k=1,2}=\frac{1}{4}\left(1+(-1)^j y \cos (\theta )+(-1)^k\sqrt{t_1^2 \sin ^2(\theta )
  +\left(x+(-1)^j t_3 \cos (\theta )\right){}^2}\right).
\end{eqnarray}

The derivative (\ref{cth}) is zero for either $\sin\theta=0$ or $H_{\theta}=0$.
$\sin\theta=0$ gives an extreme point $\theta=\theta_e=0$.
While $H_{\theta}=0$ has the one obvious solution $\theta_e=\pi/2$ and a special solution $\theta_s$ that
depends on the density matrix entries.
The optimization problem is then reduced to study the second derivative of $G(\theta, 0)$ with respect to $\theta$
evaluated at the critical angles $\theta_e=0,\pi/2$ and $\theta_s$. Denote $H_{\theta}'=\partial H_{\theta}/\partial \theta$.
The second derivative $\partial^2 G/\partial^2 \theta
=-(\cos\theta\, H_{\theta}+\sin\theta\, H_{\theta}')/4$ evaluated at the three $\theta_e$s depends on the
behavior of two quantities,
\begin{eqnarray}\label{cri1}
  H_0=&&-\partial^2 G/\partial^2 \theta|_{\theta=0}\nonumber\\
  =&&\frac{t_1^2-t_3
|x+t_3|}{|x+t_3|}\log_2\frac{p_+ +|x+t_3|}{p_+ -|x+t_3|}+y\log_2\frac{p_-^2-(x-t_3)^2}{p_+^2-(x+t_3)^2}\nonumber\\[2mm]
&&+\frac{t_1^2+t_3
|x-t_3|}{|x-t_3|}\log_2\frac{p_- +|x-t_3|}{p_- -|x-t_3|},
\end{eqnarray}
and
\begin{eqnarray}\label{cri2}
H_{\pi/2}'=&&-\partial^2 G/\partial^2 \theta|_{\theta=\pi/2}\nonumber\\
=&&-4
[(t_3^2 x^2+(y^2-2 t_3 x y )(t_1^2+x^2))\sqrt{t_1^2+x^2}\nonumber\\
+&&
t_1^2 \left(t_1^2+x^2-1\right) \left(t_1^2-t_3^2+x^2\right) \tanh ^{-1}\left(\sqrt{t_1^2+x^2}\right)]/[(t_1^2+x^2-1) (t_1^2+x^2){}^{3/2}].
\end{eqnarray}
The sign of $H_0$ and $H_{\pi/2}'$ determines which of $G(0,0)$, $G(\pi/2, 0)$ and $G(\theta_s, 0)$ is the minimum.

(1) If $H_0>0$ and $H_{\pi/2}'>0$, then $\theta_s$ takes values in $(0, \pi/2)$. In this case, the minimum of
$G(\theta_s, 0)$ depends on the state. For given $\rho_{AB}$, $\theta_s$ can be calculated numerically from $H_\theta=0$.

(2) Otherwise, we have the minimum either $G(0,0)$ or $G(\pi/2, 0)$,
\begin{eqnarray}\label{g0}
G(0,0)=-\sum_{k,j\in\{0,1\}} Q_{k,j}\log_2 Q_{k,j},
\end{eqnarray}
with $Q_{k,j}=\frac14[(1+(-1)^kx)+(-1)^j(y+(-1)^kt_3)]$, and
\begin{equation}
  G(\pi/2,0)=1+\mathcal{L}(\frac{1}{2}\left(1-\sqrt{t_1^2+x^2}\right)),
\end{equation}
where $\mathcal{L}(w)=-w\log_2 w- (1-w)\log_2(1-w)$ is the binary entropy.
Thus, we have the following result:
one-way quantum deficit of $\rho_{AB}$ is given by
\begin{equation}\label{a}
\tilde{\Delta}=\sum_\pm (u_\pm\log_2 u_\pm+ v_\pm\log_2 v_\pm)+ G,
\end{equation}
where
\begin{equation}
G=\left\{ \begin{array}{ll}
                                   G(\theta_s,0), & \hbox{$H_0>0$ and $H_{\pi/2}'>0$, $\theta_s\in(0, \pi/2)$}; \\
                                   \min\{G(0,0), G(\pi/2,0)\}, & \hbox{others}.
                                 \end{array}
                               \right.
\end{equation}

By careful numerical analysis, there is at most one zero point of first derivative of $G(\theta,0)$ with respect to $\theta$, and only
when $H_0>0$ and $H_{\pi/2}'>0$, one gets the minimum inside the interval $\theta_s\in(0, \pi/2)$.
Therefore, we can obtain the analytical minimum of $G$ at $\theta=0, \pi/2$ or $\theta=\theta_s$.
In the following we present some detailed examples.

{\sf Example 1}. Consider the class of special $X$ states defined by
\begin{equation}
\rho_{AB}=q|\psi^-\rangle\langle\psi^-|+(1-q)|00\rangle\langle00|,
\end{equation}
where $|\psi^-\rangle=\frac1{\sqrt{2}}(|01\rangle-|10\rangle)$, $q\in[0,1]$.

By using the Bloch sphere representation, from (\ref{cri1}) and (\ref{cri2}) we have
\begin{eqnarray}
H_0(q)=\frac{(q-1)}{\left| 2-3 q\right| } \left[\left| 2-3 q\right|
\log_2 \left(\frac{2}{q}-2\right)+(5 q-2) \log_2
\left(\frac{q -2+\left| 2-3 q\right|}{q-2-\left| 2-3 q\right| }\right)\right]
\end{eqnarray}
and
\begin{eqnarray}
H_{\pi/2}'=\frac{4 (1-q)}{\left(2 q^2-2 q+1\right)^{3/2}} \left(\sqrt{2 q^2-2 q+1} \left(4 q^2-3 q+1\right)-2 q^3 \tanh ^{-1}\left(\sqrt{2 q^2-2 q+1}\right)\right).
\end{eqnarray}

\begin{figure}[h] \centering
\subfigure[] { \label{q}
\includegraphics[width=0.45\columnwidth]{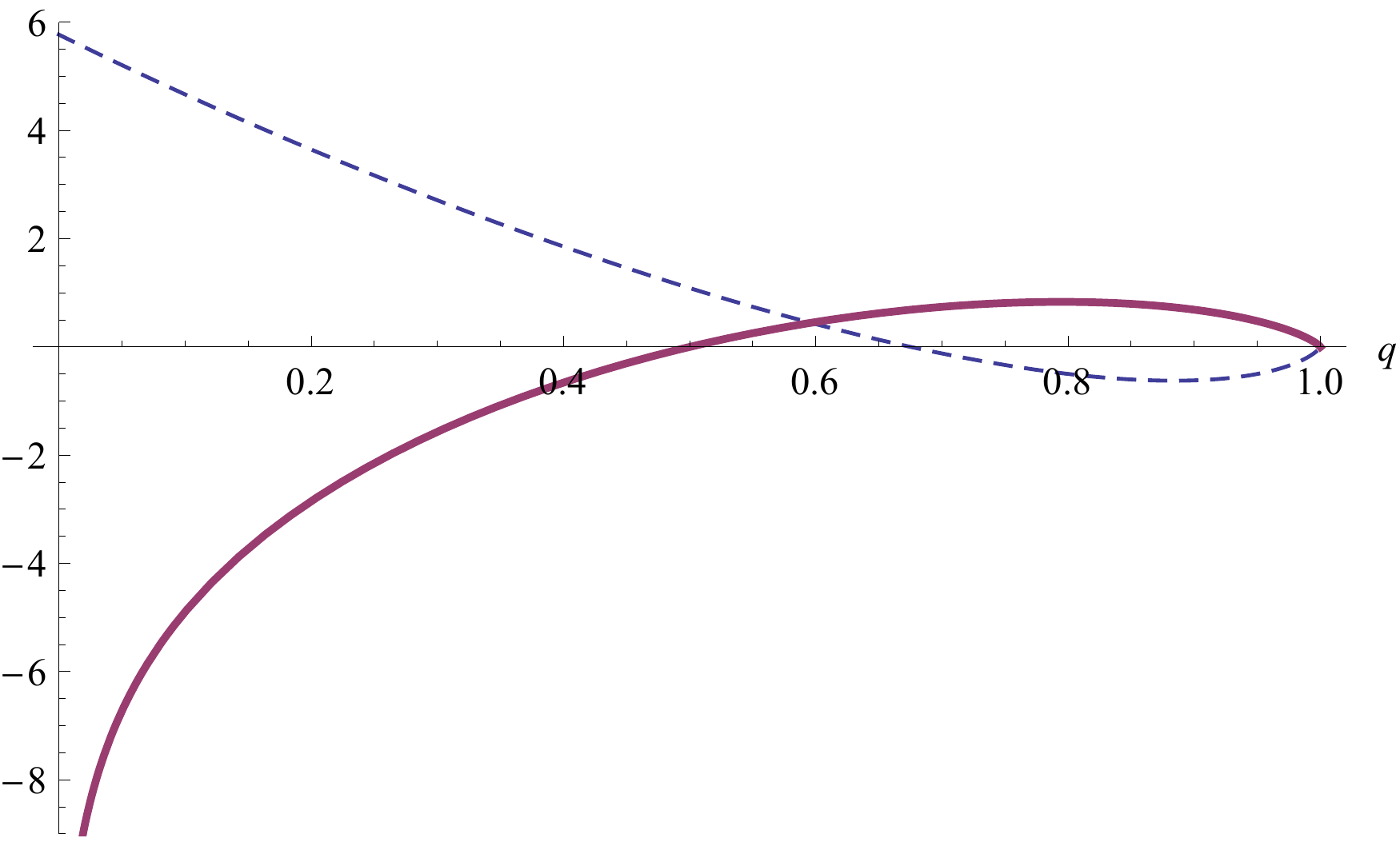}
}
\subfigure[] { \label{qa}
\includegraphics[width=0.45\columnwidth]{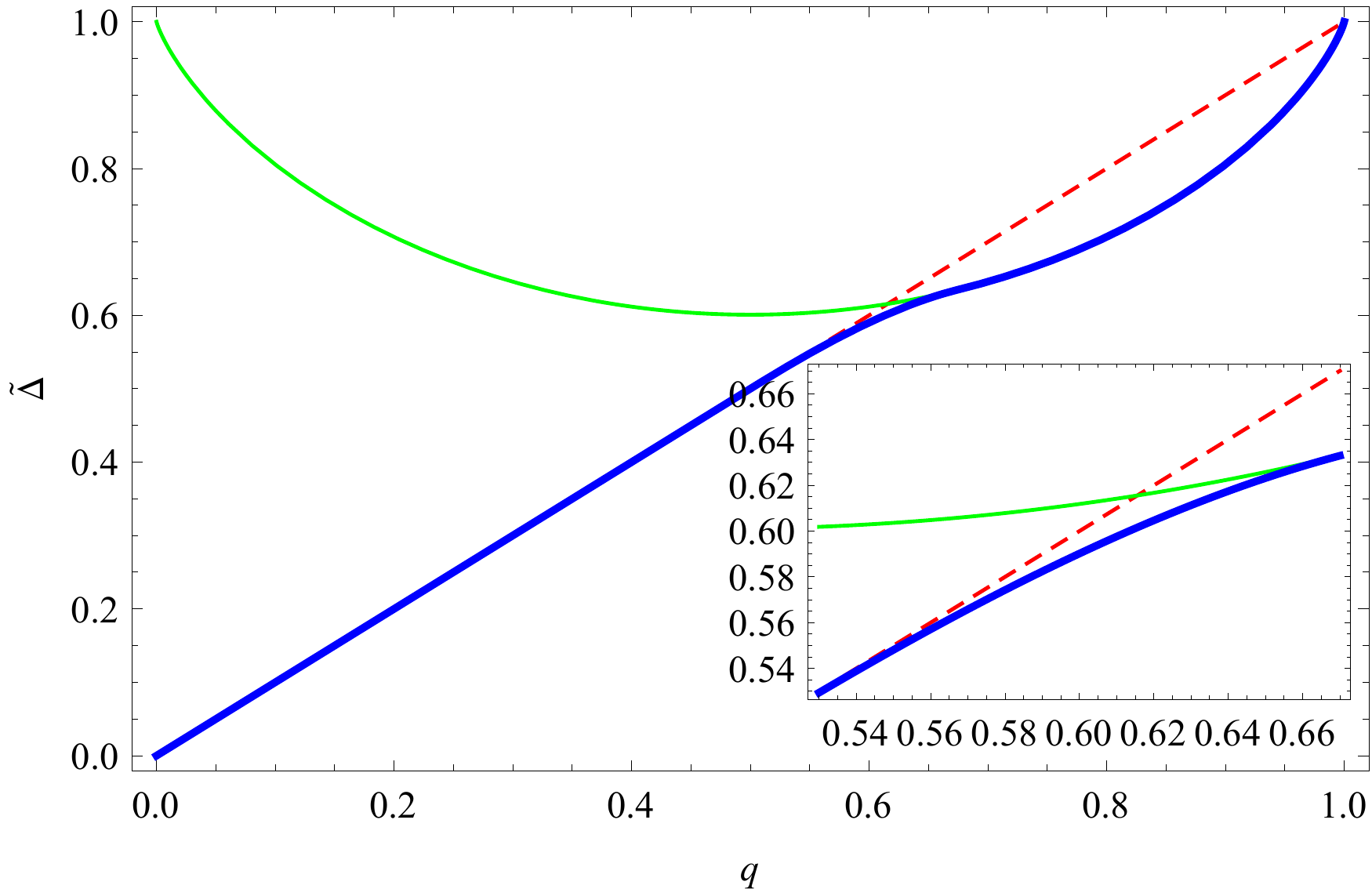}
}
\caption{(a) $H_0$ (purple thick solid line) and
$H_{\pi/2}'$ (blue dashed line) with respect to $q$. (b) One-way quantum deficit (blue thick line) via $q$.
The dashed red line stands for $G(0,0)$ which is the one-way quantum deficit for
$q\in[0,1/2]$. The green line is for $G(\pi/2,0)$. It coincides with the one-way quantum deficit
for $q\in(0.67,1$].}\label{m}
\end{figure}

The case that both $H_0(q)>0$ and $H_{\pi/2}'(q)>0$ happens only in interval $q\in (1/2, 0.67)$, see Fig.\ref{q}, where
$q=1/2$ and $0.67$ are the solutions of $H_0(q)=0$ and $H_{\pi/2}'=0$, respectively.
From (\ref{a}), we have
\begin{eqnarray}
\tilde{\Delta}=\left\{
                   \begin{array}{ll}
                     q, &~~~ \hbox{$\theta=0$, $q\in[0,1/2]$;} \\
                     G(\theta_s,0), &~~~ \hbox{$q\in (1/2,0.67]$;}\\[1mm]
q \log_2q+(1-q) \log_2 (1-q)+1+\mathcal{L}(\frac{\left(1+\sqrt{q^2+(1-q)^2}\right)}{2}),
&~~~ \hbox{$\theta=\pi/2$, $q\in(0.67,1$],}
                   \end{array}
                 \right.
\end{eqnarray}
see Fig.\ref{qa}.

{\sf Example 2}. We consider the state
\begin{eqnarray}
  \rho_\alpha&&=\alpha|\phi\rangle\langle\phi|+(1-\alpha)/2(|01\rangle\langle01|+|10\rangle\langle10|),
\end{eqnarray}
with $|\phi^+\rangle=(|00\rangle+|11\rangle)/\sqrt{2}$.

\begin{figure}[htpb] \centering
\subfigure[] { \label{c2}
\includegraphics[width=0.5\columnwidth]{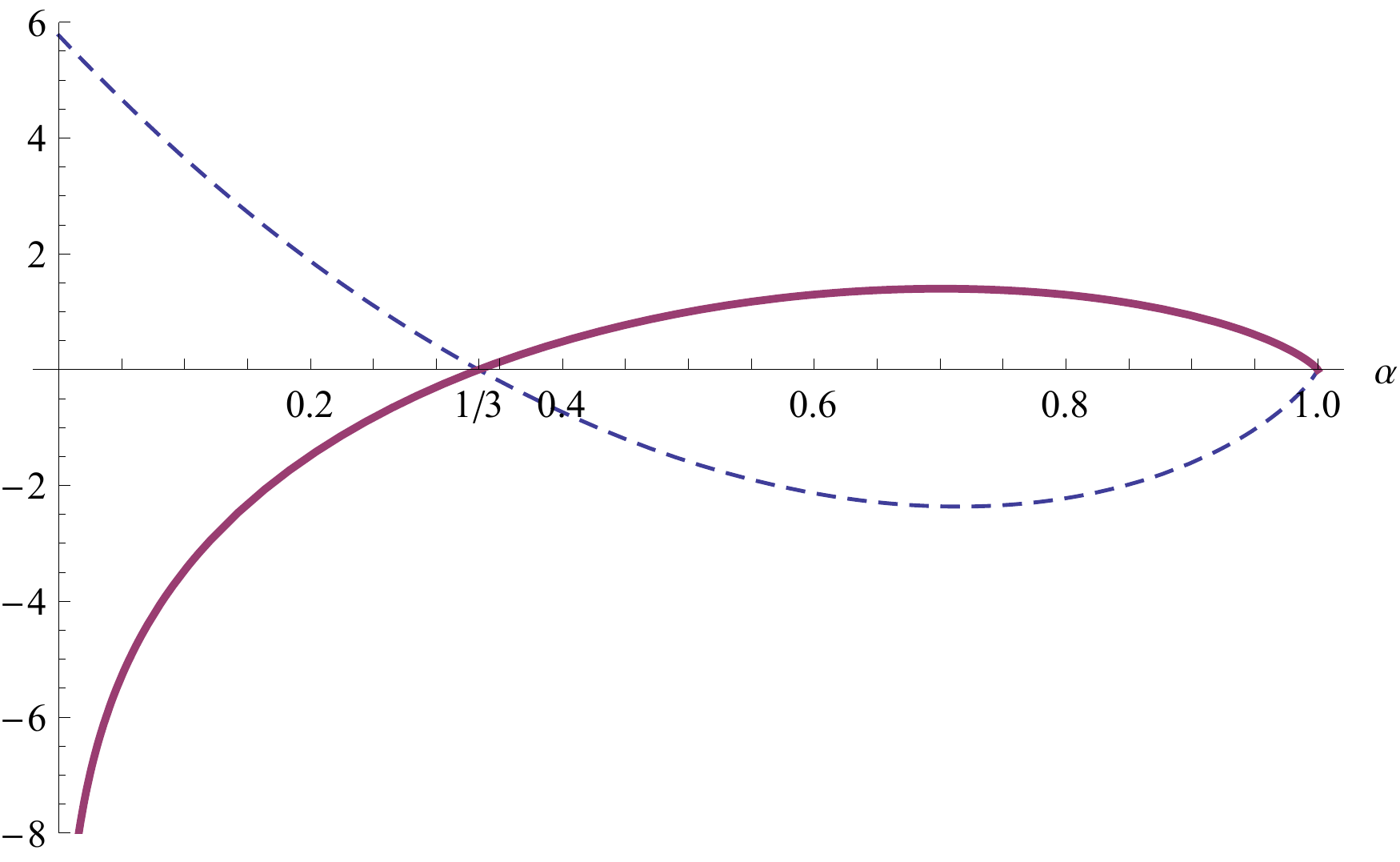}
}%
\subfigure[] { \label{a2}
\includegraphics[width=0.5\columnwidth]{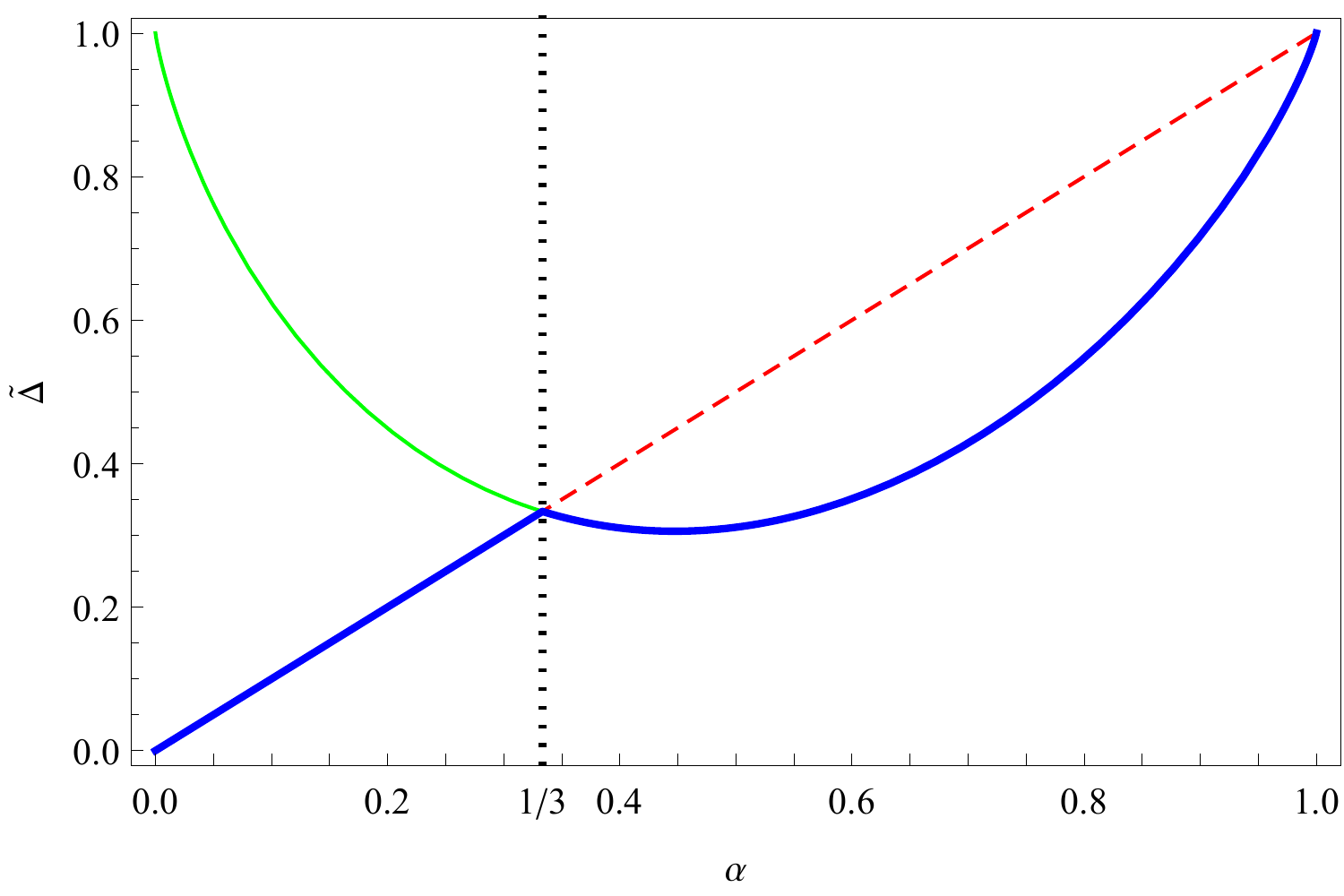}
}
\caption{ (a) $H_0$ (purple thick solid line) and
$H_{\pi/2}'$ (blue dashed line) with respect to $\alpha$. (b)
One-way quantum deficit (blue thick line) via $\alpha$. Dashed red line for $G$ at
$\theta=0$, and green line for $G$ at $\theta=\pi/2$.}\label{astate}
\end{figure}

For this case we have
\begin{eqnarray}
  H_0(\alpha)=\frac{2 (\alpha -1) (3 \alpha -1)
\left(\log_2 \left(1-|2 \alpha -1|\right)
-\log_2 \left(1+|2 \alpha -1|\right)\right)}{|2 \alpha -1|},
\end{eqnarray}
and
\begin{eqnarray}
  H_{\pi/2}'=4 (\alpha -1) (3 \alpha -1) \tanh ^{-1}(\alpha )/\alpha.
\end{eqnarray}
For $\alpha=1/3$ both $H_0(\alpha)$ and $H_{\pi/2}'(\alpha)$ are equal to
zero. There is no domain of $\alpha$ such that both $H_0>0$ and $H_{\pi/2}'>0$, see Fig. \ref{c2}.
Hence we can easily obtain the analytical expression of one-way quantum deficit for $\rho_\alpha$,
\begin{eqnarray}\label{aea}
  \tilde{\Delta}=\left\{
                   \begin{array}{ll}
                     \alpha, & \hbox{$\theta=0$, $\alpha\in[0,1/3]$;} \\
                    1+ \alpha+\alpha\log _2 \alpha
                    +\frac{1}{2} \left[(1-\alpha ) \log _2(1-\alpha )
                    -(1+\alpha)\log _2(1+\alpha)\right],
& \hbox{$\theta=\pi/2$, $\alpha\in(1/3,1]$,}
                   \end{array}
                 \right.
\end{eqnarray}
see Fig.\ref{a2} for the analytical expression of one-way quantum deficit vs $\alpha$.

{\sf Example 3}. Consider the Bell diagonal state,
\begin{eqnarray}
  \rho_{AB}=\frac14[I\otimes I +\sum_{i=1}^3t_i\sigma_i\otimes\sigma_i].
\end{eqnarray}
We have
\begin{eqnarray}
  H_0=\frac{2 \left(t_1^2-t_3^2\right)}{|t_3|}
\log_2 \left(\frac{ 1+|t_3|}{1-|t_3|}\right)
\end{eqnarray}
and
\begin{equation}
  H_{\pi/2}'=-4 \left(t_1^2-t_3^2\right) \tanh ^{-1}\left(|t_1|\right)/|t_1|.
\end{equation}
Since $\log_2 \frac{ 1+|t_3|}{1-|t_3|}>0$ and $\tanh ^{-1}\left(|t_1|\right)>0$,
$H_0$ and $H_{\pi/2}'$ cannot be greater than zero simultaneously. We obtain the
analytical expression $\min \{G(0, 0), G(\pi/2, 0)\}$,
\begin{eqnarray}
  G(0,0)=1+\mathcal{L}(\frac{1+|t_3|}{2})
\end{eqnarray}
and
\begin{eqnarray}
  G(\pi/2,0)=1+\mathcal{L}(\frac{1+|t_1|}{2}).
\end{eqnarray}
Therefore $\min\{G(0,0), G(\pi/2,0)\}=1+\mathcal{L}(\frac{1+t}{2})$, where $t=\max\{|t_1|,|t_3|\}$.
In fact, for Bell-diagonal states, the optimization is ontained at $t_3=\pm t_1$ or $t_3=\pm t_2$ \cite{chen11}.
Therefore, we recovered the result in Ref.\cite{wyk} where $t=\max\{|t_1|,|t_2|,|t_3|\}$.

%
%

\section{One-way quantum deficit under decoherence}

A system undergoes environmental noises can be characterized by
Kraus operators. We consider quantum two-qubit systems subjecting to dephasing channels described by
the Kraus operators $F_0=|0\rangle\langle0|+\sqrt{1-\gamma}|1\rangle\langle1|$ and
$F_1=\sqrt{\gamma}=|1\rangle\langle1|$, where $\gamma=1-e^{-\tau t}$
and $\tau$ is phase damping rate \cite{nielsen}.
Under the channel the $\rho_{AB}$ is changed to be
\begin{eqnarray}
\rho_{AB}'&&=\sum_{i,j=0}^1F_i^A\otimes F_j^B\rho_{AB} (F_i^{A}\otimes F_j^{B})^\dagger\nonumber\\
&&=\frac14[I\otimes I +x \sigma_3\otimes I
  + y I\otimes  \sigma_3+\sum_{i=1}^2(1- \gamma)^2t_i\sigma_i\otimes\sigma_i+(t_3\sigma_3\otimes\sigma_3)].\nonumber
\end{eqnarray}
We see that $t_1$ and $t_2$ have been transformed to $(1-\gamma)^2t_1$ and $(1-\gamma)^2t_2$.
We have
\begin{eqnarray}
  H_0=&&-\partial^2 G/\partial^2 \theta|_{\theta=0}\nonumber\\
  =&&\frac{\Upsilon^2-t_3
|x+t_3|}{|x+t_3|}\log_2\frac{p_+ +|x+t_3|}{p_+ -|x+t_3|}+y\log_2\frac{p_-^2-(x-t_3)^2}{p_+^2-(x+t_3)^2}\nonumber\\
&&+\frac{\Upsilon^2+t_3
|x-t_3|}{|x-t_3|}\log_2\frac{p_- +|x-t_3|}{p_- -|x-t_3|}
\end{eqnarray}
and
\begin{eqnarray}
H_{\pi/2}'=&&\Upsilon^2 \left(\Upsilon^2+x^2-1\right) \left(\Upsilon^2-t_3^2+x^2\right) \tanh ^{-1}\left(\sqrt{\Upsilon^2+x^2}\right)]/(\left(\Upsilon^2+x^2-1\right) \left(\Upsilon^2+x^2\right){}^{3/2})\nonumber\\
&&-4
[(-2 t_3 x y (\Upsilon^2+x^2) +x^2 y^2 +\Upsilon^2 y^2 +t_3^2 x^2 )\sqrt{\Upsilon^2+x^2},
\end{eqnarray}
where $\Upsilon=(1-\gamma)^2t_1$. It is direct to verify that
$G(0,0)$ exactly given by (\ref{g0}). While $G(\pi/2,0)$ has the following form,
\begin{equation}
G(\pi/2,0)=1+\mathcal{L}(\frac{1}{2}\left(1-\sqrt{\Upsilon^2+x^2}\right)).
\end{equation}

\begin{figure}
  \includegraphics[width=8cm]{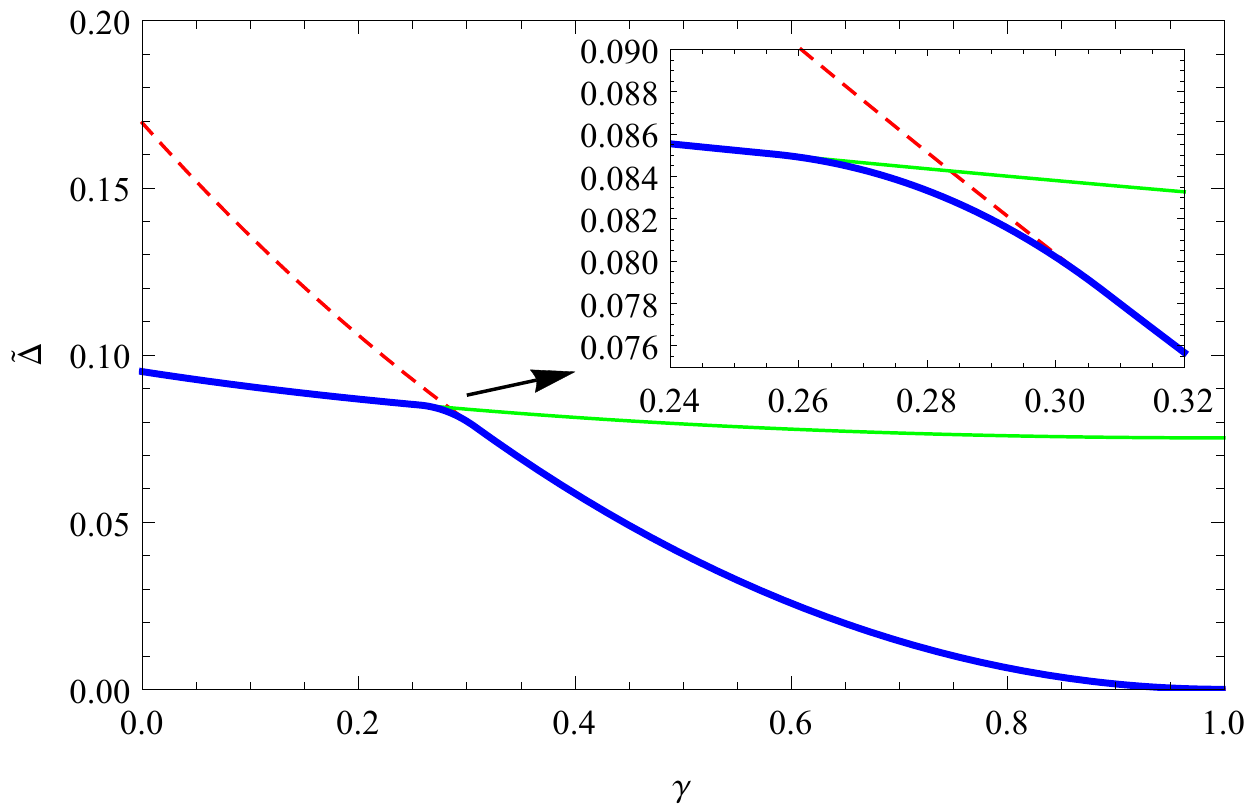}\\
  \caption{One-way quantum deficit vs $\gamma$ under dephasing noise
for $x=0.45$, $y=0.32$, $t_1=0.43$, $t_2=0.09$ and $t_3=0.15$.}\label{deph}
\end{figure}

As an example, taking $x=0.45$, $y=0.32$, $t_1=0.43$, $t_2=0.09$ and $t_3=0.15$,
we can observe the one-way quantum deficit under
phase damping channel, see Fig. (\ref{deph}).
It should be emphasized that the exact boundaries exist between three different branches and sudden transitions occur in the phase damping channel.

\section{Summary}

We have provided an effective approach to get analytical results of one-way
quantum deficit for general two-qubit $X$ states.
Analytical formulae of one-way
quantum deficit have been obtained in general for states such that
$\min\{G(0,0), G(\pi/2,0)\}$.
It has been shown that only in very few cases, the conditions
$H_0>0$ and $H_{\pi/2}'>0$ are satisfied. Even in such special cases,
the one-way quantum deficit can be easily calculated by
solving $\theta_s$ from a one-parameter equation.
We have also studied the decoherence of one-way quantum deficit under phase damping channel.
As for a detailed example, it has been shown that the one-way quantum deficit changes gradually
phase damping channel. There is no behavior like sudden death of quantum entanglement.

\vspace{2.5ex}
\noindent{\bf Acknowledgments}\, \,
We thank the anonymous referee for the useful suggestions and valuable comments. This work is supported by the NSFC under number 11275131.


\begin{thebibliography}{99}

\bibitem{hhhh} Amico, L., Fazio, R., Osterloh, A., Vedral, V.: Rev. Mod. Phys. {\bf 80}, 517 (2008)

\bibitem{horodecki09}Horodecki, R., Horodecki, P., Horodecki, M.,
 Horodecki, K.: Rev. Mod. Phys.
    \textbf{81}, 865 (2009)

\bibitem{modi12} Modi, K.,  Brodutch, A., Cable, H., Paterek, T.,
   Vedral, V.:  Rev. Mod. Phys. \textbf{84}, 1655 (2012)

\bibitem{datta08}
Datta, A., Shaji, A., Caves, C. M.: {Phys. Rev. Lett.} {\bf 100}, 050502 (2008)

\bibitem{lanyon08}
Lanyon, B. P., Barbieri, M., Almeida, M. P., White, A. G.: {Phys. Rev.
Lett.}, {\bf 101}, 200501 (2008)

\bibitem{zurek01} Ollivier, H., Zurek, W. H.: Phys. Rev.
    Lett. \textbf{88}, 017901(2001)

\bibitem{henderson01} Henderson, L.,  Vedral, V.:
    J. Phys. A \textbf{34}, 6899 (2001)

\bibitem{roa11}
Roa, L.,  Retamal, J. C., Alid-Vaccarezza, M.:  Phys. Rev. Lett. {\bf 107}, 080401 (2011)

\bibitem{libo12}
Li, B., Fei, S. M., Wang, Z. X., Fan, H.: {Phys. Rev.} A {\bf 85}, 022328 (2012)

\bibitem{da12}
Daki¡äc, B.,  et al.: Nat. Phys. {\bf 8}, 666 (2012)

\bibitem{maca}
Madhok, V., Datta, A.: Phys. Rev. A {\bf 83}, 032323 (2011)

\bibitem{cavalcanti11}
Cavalcanti, D., Aolita, L., Boixo, S., Modi, K., Piani, M., Winter, A.: Phys. Rev. A {\bf 83}, 032324 (2011)

\bibitem{luo2}
Luo, S.: Phys. Rev. A {\bf 77}, 022301 (2008)

\bibitem{daluo}
Daki\'c, B., Vedral, V., Brukner, C.:  Phys. Rev. Lett. {\bf 105}, 190502 (2010)


\bibitem{luo10}
Luo, S., Fu, S.:  Phys. Rev. A {\bf 82}, 034302 (2010)

\bibitem{modi}
Modi, K.,  Paterek, T., Son, W., Vedral, V., Williamson, M.:  Phys. Rev. Lett. {\bf 104}, 080501 (2010)

\bibitem{adesso}
Adesso, G., Datta, A.:  Phys. Rev. Lett. {\bf 105}, 030501 (2010)

\bibitem{giorda}
Giorda, P., Paris, M. G. A.:  Phys. Rev. Lett. {\bf 105}, 020503 (2010)

\bibitem{luo08} Luo, S.: Phys. Rev. A \textbf{77}, 042303 (2008)

\bibitem{ali10}
Ali, M., Rau, A. R. P., Alber, G.: Phys. Rev. A {\bf 81}, 042105
 (2010); Erratum in: Phys. Rev. A {\bf 82}, 069902(E) (2010)

\bibitem{lu11}
Lu, X.-M., Ma, J., Xi, Z., Wang, X.: Phys. Rev. A {\bf 83}, 012327 (2011)

\bibitem{chen11}
Chen, Q., Zhang, C., Yu, S., Yi, X. X., Oh, C. H.: Phys. Rev. A
 {\bf 84}, 042313 (2011)

\bibitem{davi}
Girolami, D., Adesso, G.: Phys. Rev. A {\bf 83}, 052108 (2011)


\bibitem{huang13}
Huang, Y.: Phys. Rev. A {\bf 88}, 014302 (2013)

\bibitem{loa}
Maldonado-Trapp, A., Hu, A., Roa, L.: Quantum Inf. Process. {\bf 14}, 1947 (2015)

\bibitem{yur}
Yurishchev, M. A.: J. Exp. Theor. Phys. {\bf 119}, 828 (2014)

\bibitem{yur15}
Yurischev, M. A.: Quantum Inf. Process. {\bf 14}, 3399 (2015)


\bibitem{opp02}
Oppenheim, J., Horodecki, M., Horodecki, P., Horodecki, R.: Phys. Rev. Lett. {\bf 89}, 180402 (2002)


\bibitem{horo05} Horodecki, M., Horodecki, P., Horodecki, R., Oppenheim, J., Sen(De), A., Sen, U., Synak, B.:  Phys. Rev. A {\bf 71}, 062307 (2005)

\bibitem{oppenheim2} Oppenheim, J., Horodecki, K., Horodecki, M.,  Horodecki, P., Horodecki, R.:  Phys. Rev. A {\bf 68}, 022307 (2003)

\bibitem{wyk}
Wang, Y. K., Ma, T., Li, B., Wang, Z. X.:  Commun. Theor. Phys. \textbf{59}, 540 (2013)

\bibitem{wyk14}
Wang, Y. K., Jing, N. H., Fei, S. M., Wang, Z. X., Cao, J. P., Fan, H.: Quantum Inf. Process. {\bf 14}, 2487 (2015)

\bibitem{str11}
Streltsov, A., Kampermann, H., Bruss, D.:  Phys. Rev. Lett. {\bf 106}, 160401 (2011)

\bibitem{zurek03}
Zurek, W. H.:  Phys. Rev. A {\bf 67}, 012320 (2003)

\bibitem{shi}
Shi, M., Sun, C., Jiang, F., Yan, X., Du, J.:
{Phys. Rev.} A {\bf 85}, 064104 (2012)

\bibitem{galve11}
Galve, F., Giorgi, G., Zambrini, R.: Europhys. Lett. \textbf{96}, 40005 (2011)

\bibitem{nielsen} Nielsen, M. A., Chuang, I. L.:
    \emph{Quantum Computation and Quantum Information},
    (Cambridge University Press,  Cambridge,  UK,  2000)
\end{thebibliography}
\end{document}